\documentclass[graybox,natbib,10pt]{svmult}
\usepackage{mathptmx}
\usepackage{amsmath}
\usepackage{helvet}
\usepackage{courier}
\usepackage{type1cm}
\usepackage{makeidx}
\usepackage{graphicx}
\usepackage[bottom]{footmisc}
\makeindex

\def\ltsima{$\; \buildrel < \over \sim \;$}    
\def\lesssim{\lower.5ex\hbox{\ltsima}}           
\def\gtsima{$\; \buildrel > \over \sim \;$}    
\def\gtrsim{\lower.5ex\hbox{\gtsima}}           

\newcommand{\aap}{A\&A}
\newcommand{\apj}{ApJ}
\newcommand{\apjl}{ApJL}
\newcommand{\apjs}{ApJS}
\newcommand{\apss}{ApSS}
\newcommand{\araa}{ARAA}
\newcommand{\mnras}{MNRAS}
\newcommand{\physrep}{Phys. Rep.}
\newcommand{\pre}{Phys.\ Rev.\ E}

\newcommand{\barr}{\begin{eqnarray}}
\newcommand{\beq}{\begin{equation}}
\newcommand{\cs} {c_{\rm s}}
\newcommand{\earr}{\end{eqnarray}}
\newcommand{\eeq}{\end{equation}}
\newcommand{\gamef} {{\gamma_{\rm e}}}
\newcommand{\kms}{~{\rm km~s}^{-1}}

\newcommand{\mh} {m_{\rm H}}
\newcommand{\Mmol}{M_{\rm mol}}
\newcommand{\Ms}{M_{\rm s}}
\newcommand{\Msun}{M_\odot}
\newcommand{\pcc} {{\rm ~cm}^{-3}}
\newcommand{\Peq} {P_{\rm eq}}

\newcommand{\Rey} {R_{\rm e}}
\newcommand{\tcool} {\tau_{\rm c}}
\newcommand{\tff} {\tau_{\rm ff}}
\newcommand{\tturb} {\tau_{\rm t}}
\newcommand{\uu}{{\bf u}}

\begin{document}

\title*{Physical Processes of Interstellar Turbulence}
\titlerunning{Physical Processes of Interstellar Turbulence}
\author{{\bf Enrique V\'azquez-Semadeni$^1$}
\\ \vspace{1\baselineskip}
$^1$Centro de Radioastronom\'ia y Astrof\'isica, UNAM Campus Morelia,
58089 M\'exico. 
E-mail: e.vazquez@crya.unam.mx
}
\authorrunning{E.\ V\'azquez-Semadeni}

\thispagestyle{empty}
\maketitle
\thispagestyle{empty}
\setcounter{page}{1}

\abstract{
This review discusses the role of radiative heating and cooling, as well
as self-gravity, in shaping the nature of the turbulence in the
interstellar medium (ISM) of our galaxy. The ability of the gas to
radiatively cool, while simultaneously being immersed in a radiative
heat bath, causes it to be much more compressible than if it were
adiabatic, and, in some regimes of density of density and temperature,
to become thermally unstable, and thus tend to spontaneously
segregate into separate phases, one warm and diffuse, the other dense
and cold. On the other hand, turbulence is an inherently mixing process,
thus tending to replenish the density and temperature ranges that would
be forbidden under thermal processes alone. The turbulence
in the ionized ISM appears to be transonic (i.e, with Mach numbers $\Ms
\sim 1$), and thus to behave essentially incompressibly. However, in the
neutral medium, thermal instability causes the sound speed of the gas to
fluctuate by up to factors of $\sim 30$, and thus the flow can be highly
supersonic with respect to the dense, cold gas, although numerical
simulations suggest that the supersonic velocity dispersion corresponds
more to the ensemble of cold clumps than to the clumps' internal
velocity dispersion. Finally, coherent large-scale compressions in the
warm neutral medium (induced by, say, the passage of spiral arms or by
supernova shock waves) can produce large, dense clouds that are affected
by their own self-gravity, and begin to contract gravitationally.
Because they are populated by large-amplitude density fluctuations,
whose local free-fall times can be significantly smaller than that of
the whole cloud, the fluctuations terminate their collapse earlier,
giving rise to a regime of hierarchical gravitational fragmentation,
with small-scale collapses occurring within larger-scale ones. Thus, the
``turbulence'' in the cold, dense clouds may be dominated by a
gravitationally contracting component at all scales.}

\section{Introduction}
\label{sec:intro}

Our galaxy, the Milky Way (or simply, the Galaxy) is a flattened
conglomerate of stars, gas, dust, and other debris, such as planets,
meteorites, etc., with a total mass $\sim 10^{12} \Msun$, where $\Msun =
2 \times 10^{33}$ g is the mass of the Sun. Most of this mass is
believed to be in a roughly spherical dark matter halo, while $\sim
10^{11} \Msun$ are in stars, and $\sim 10^{10} \Msun$ are contained in
the gaseous component, mostly confined to the Galactic disk
\citep{AAQ00}.

The gasesous component may be in either ionized, neutral
atomic or neutral molecular forms, spanning a huge range of densities
and temperatures, from the so-called hot ionized medium (HIM), with
densities $n \sim 10^{-2} \pcc$ and temperatures $T \sim 10^6$ K, through
the warm ionized and neutral (atomic) media (WIM and WNM, respectively),
with $n \sim 0.3 \pcc$ and $T \sim 10^4$ K and the cold neutral (atomic)
medium (CNM, $n \sim 30 \pcc$, $T \sim 100$ K, to the {\it giant
molecular clouds} (GMCs, $n \gtrsim 100 \pcc$ and $T \sim 10$--20
K. GMCs can span several tens of parsecs across, and, in turn, have
plenty of substructure, which are commonly referred to as {\it clouds}
($n \sim 10^3 \pcc$, size scales $L$ of a few parsecs), {\it clumps} ($n
\sim 10^4 \pcc$, $L \sim 1$ pc), and {\it cores} ($n \gtrsim 10^5 \pcc$,
$L \sim 0.1$ pc), although the temperature of all molecular gas is
remarkably uniform, $\sim 10$--30 K \citep[e.g.,][]{Ferriere01}. 

The gaseous component, along with the dust, a cosmic-ray background, and
a magnetic field of mean intensity of a few $\mu$G constitute what we
know as the {\it interstellar medium} (ISM). This medium is in most
cases well described by the fluid approximation \citep[][Ch.\
1]{Shu92}. Moreover, the ISM is most certainly turbulent, as typical
Reynolds numbers in it are very large. For example, in the cold ISM,
$R_{\rm e} \sim 10^5$--$10^7$ \citep[][Sec.\ 4.1]{ES04}. This is mostly
due to the very large spatial scales involved in interstellar flows.
Because the ISM's temperature varies so much from one type of region to
another, so does the sound speed, and the flow is often super- or
trans-sonic \citep[e.g., ][and references therein]{HT03, ES04}. This
implies that the flow is significantly compressible, inducing
significant density fluctuations (Sec.\ \ref{sec:Re_Ms}).

In addition to being turbulent, the ISM is subject to a number of
additional physical processes, such as gravitational forces exerted by
the stellar and dark matter components as well as by the ISM itself
(i.e., by its {\it self-gravity}), magnetic fields, cooling by radiative
microscopic processes, and radiative heating due both to nearby stellar
sources as well as to diffuse background radiative fields. All of this
adds up to make the ISM an extremely complex and dynamical medium.

Finally, the self-gravity of the gas causes local (i.e., spatially
intermittent) gravitational collapse events, in which a certain gas
parcel within a molecular cloud goes out of equilibrium between its
self-gravity and all other forces that oppose it, undergoing an
implosion that leads the gas density to increase by many orders of
magnitude, and whose end product is a star or a group (``cluster'') of
stars.

In this review, we focus on the interaction between turbulence and the
effects of radiative heating and cooling, which effectively enhance the
compressibility of the flow, and have a direct effect on the star
formation process. The plan of the paper is as follows: in Sec.\
\ref{sec:thermodynamics} we first review the effects that the net
heating and cooling have on the effective equation of state of the flow
and, in the case of thermally unstable flows, on its tendency to
spontaneously segregate in distinct phases. Next, in Sec.\
\ref{sec:turb} we discuss a few basic notions about turbulence and the
turbulent production of density fluctuations in the compressible case,
to then discuss, in Sec.\ \ref{sec:turb_therm}, the interplay between
turbulence and the heating and cooling. In Sec.\ \ref{sec:turb_diff}, we
discuss the likely nature of turbulence in the diffuse (warm and hot)
parts of the ISM, while in Sec.\ \ref{sec:turb_dense} we do the same for
the dense, cold atomic and molecular clouds. We conclude in Sec.\
\ref{sec:conclusions} with a summary and some final remarks. Due to
space limitations, we do not discuss magnetic fields, although we refer
the interested reader to the reviews by \citet{VS+00b}, \citet{Cho+03},
\citet{ES04}, and \citet{MO07}.

\section{ISM Thermodynamics: Thermal Instability} \label{sec:thermodynamics}

The ISM extends essentially over the entire disk of the Galaxy and, when
considering a certain subregion of it, such as a cloud or cloud complex,
it is necessary to realize that any such subregion constitutes an open
system, whose interactions with its environment need to be taken into
account. A fundamental form of interaction with the surroundings, besides
dynamical interactions, is through the exchange of heat. Indeed, the ISM
is permeated by a radiation field, due to the combined shine of the
stars in the disk. Moreover, a bath of relativistic charged particles,
mostly protons, known as {\it cosmic rays}, also exists in the
ISM. These are believed to be accelerated in strong shocks produced by
supernova explosions \citep{BE87}. Both UV radiation and cosmic rays
provide heating and ionization sources for the ISM at large \citep[see,
e.g.,][]{DM72, Wolfire+95}. Finally, violent events, such as
supernova explosions, can locally heat their surroundings to very high
temperatures, causing local bubbles of hot, million-degree ionized gas.

On the other hand, the ISM can cool by emission from ions, atoms
and molecules, and by thermal emission from dust grains \citep{DM72,
SD93, Wolfire+95}. The balance between these radiative heating and
cooling radiative processes, along with the heat due to mechanical work
and thermal conductivity, determine the thermodynamic properties of the
ISM. These properties will be the focus of the present section,
rather than the detailed microphysical processes that mediate the
radiative transfer, about which there exist plenty of excellent books
and reviews \citep[e.g.,][]{DM72, SD93, OF06}.

\begin{figure}
\begin{center}
\includegraphics[scale=.27]{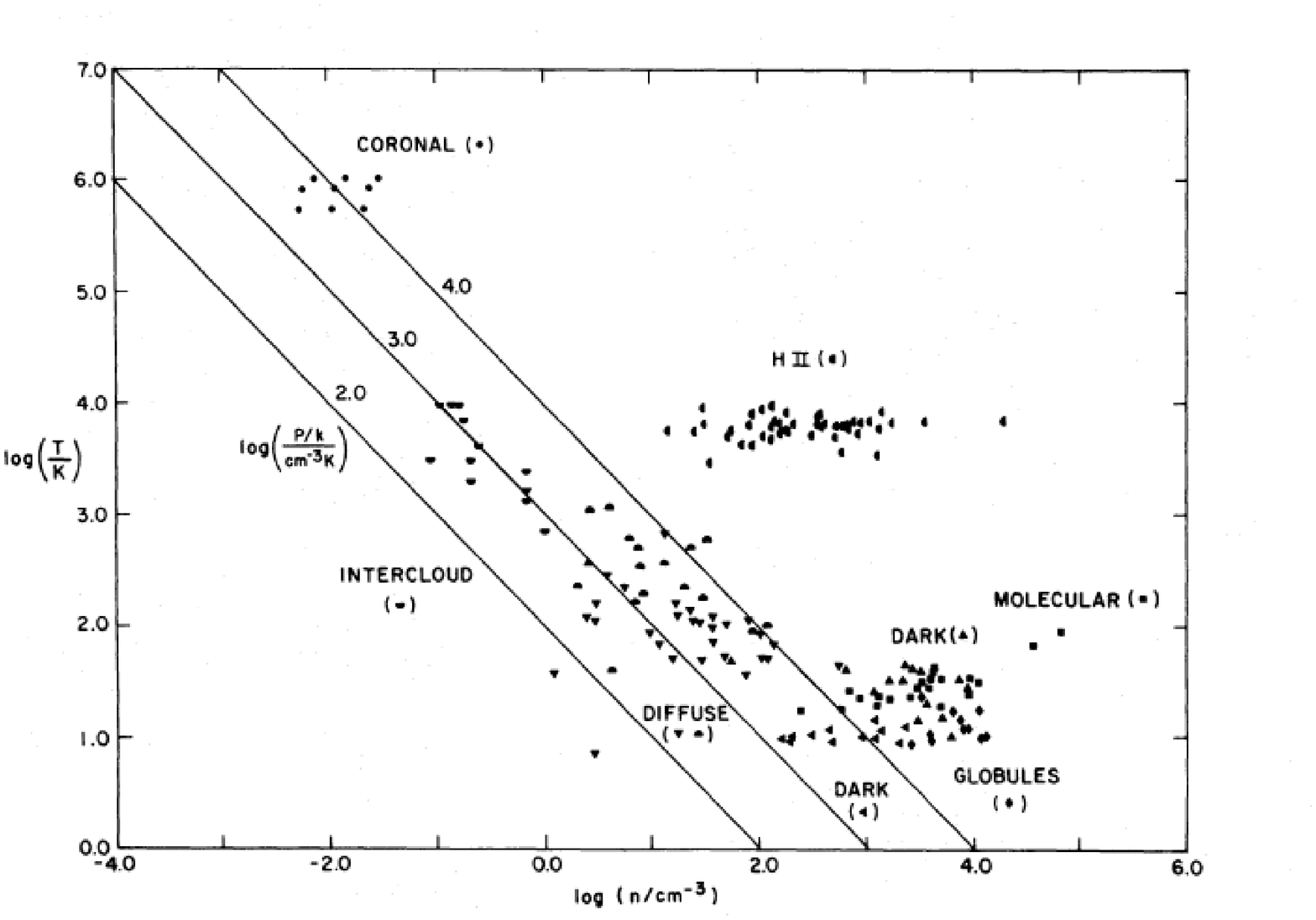}
\includegraphics[scale=.36]{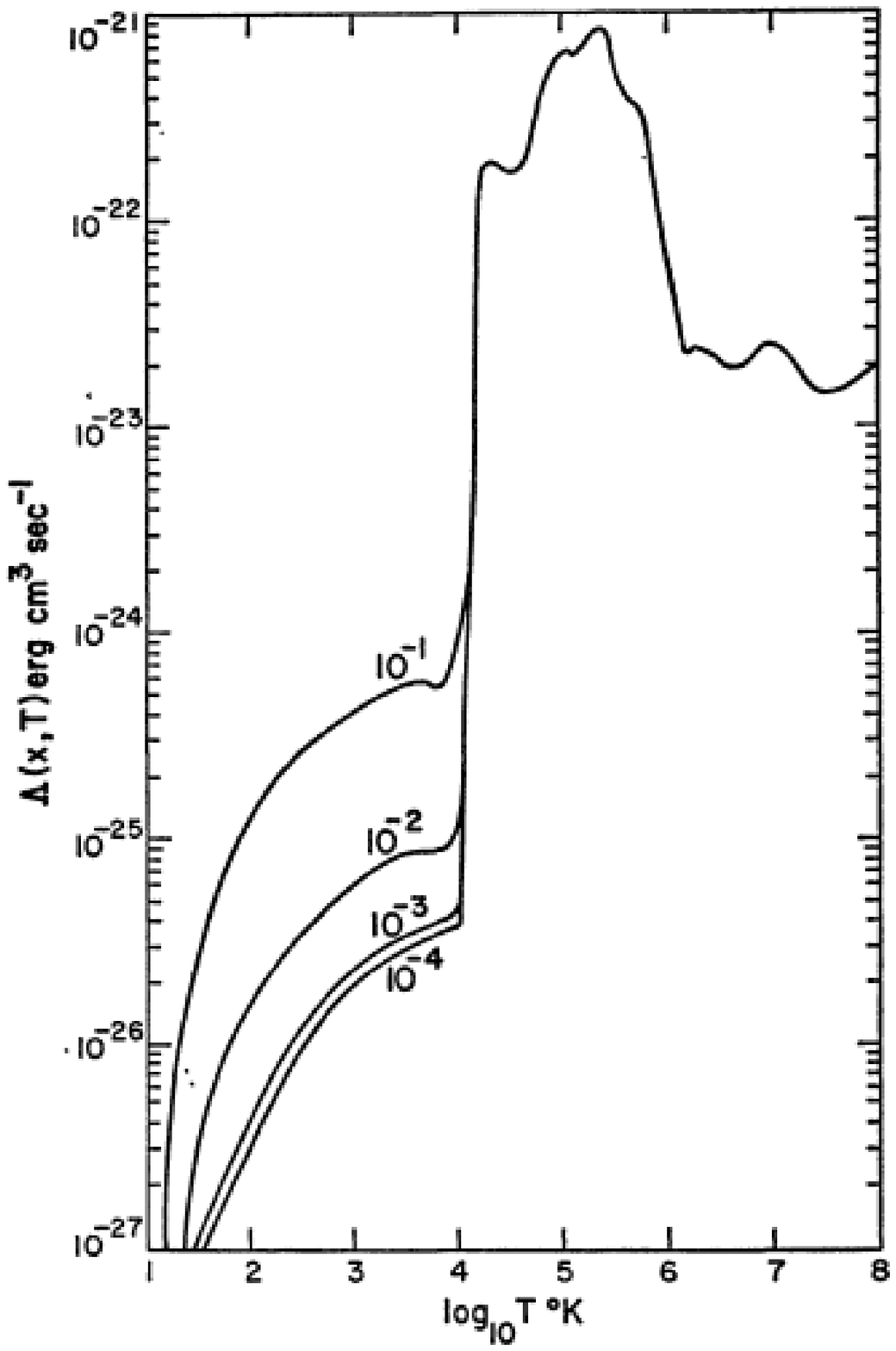}
\end{center}
\caption{{\it Left:} Thermal pressure in various types of interstellar
regions. The points labeled {\it coronal} correspond essentially to what
is referred to as the HIM in the text; {\it intercloud} regions refer to
the WIM and WNM; {\it diffuse} to CNM clouds, and {\it dark, globule}
{\it molecular} to molecular gas. From \citet{Myers78}. {\it Right:}
Temperature dependence of the cooling function. The labels indicate
values of the ionization fraction (per number) of the gas. From
\citet{DM72}.}
\label{fig:Myers78_DM72}
\end{figure}

Globally, and as a first approximation, the ISM is roughly isobaric, as
illustrated in the left panel of Fig.\ \ref{fig:Myers78_DM72}. As can be
seen there, most types of regions, either dilute or dense, lie within an
order of magnitude from a thermal pressure $P \sim 3000 $
K$\pcc$.\footnote{It is customary in Astrophysics to express pressure in
units of [K$\pcc$]. Strictly speaking, this corresponds to $P/k$, where
$k$ is the Boltzmann constant.} The largest deviations from this
pressure uniformity are found in HII regions, which are the ionized
regions around massive stars due to the star's UV radiation, and
molecular clouds, which, as we shall see in Sec.\ \ref{sec:turb_dense},
are probably pressurized by gravitational compression.

The peculiar thermodynamic behavior of the ISM is due to the functional
forms of the radiative heating and cooling functions acting on it, which
depend on the density, temperature, and chemical composition of the gas.
The right panel of Fig.\ \ref{fig:Myers78_DM72} shows the temperature
dependence of the cooling function $\Lambda$ \citep{DM72}. 

To understand the effect of the functional form of the radiative heating
and cooling functions on the net behavior of the gas, let us write the
conservation equation for the internal energy per unit mass, $e$. We
have \citep[e.g.,][]{Shu92}
\beq
\frac{\partial e}{\partial t} + \uu \cdot \nabla e = -(\gamma - 1) e
\nabla \cdot \uu  + \Gamma - n \Lambda,
\label{eq:int_en_cons}
\eeq
where we have neglected thermal conduction and heating by magnetic
reconnection. In eq.\ (\ref{eq:int_en_cons}), \uu\ is the velocity
vector, $\gamma$ is the ratio of specific heats, $\Gamma$ is the heating
rate per unit mass, $n = \rho/\mu \mh$ is the number density, with $\mu$
the mean partucle mass and $\mh$ the hydrogen atom's mass. Note that $n
\Lambda$ is the cooling rate per unit mass. The first term on the right
hand side is the $P~dV$ work per unit mass.

In the simplest possible case, that of a hydrostatic ($\uu = 0$) and 
steady ($\partial/\partial t = 0$) state, eq.\ (\ref{eq:int_en_cons})
reduces to the condition of {\it thermal equilibrium},
\beq
\Gamma(\rho,T) = n \Lambda(\rho,T),
\label{eq:therm_eq}
\eeq
where we have denoted explicitly the dependence of $\Gamma$ and
$\Lambda$ on the density and temperature, and neglected the dependence
on chemical composition. 

Let us now assume we have a gas parcel in thermal and hydrostatic
equilibrium, at a temperature somewhere in the range where the slope of
$\Lambda$ with respect to $T$ is negative $\left({\rm i.e.,~} (\partial
\Lambda / \partial T)_\rho< 0\right)$, $10^{5.5} \lesssim T
\lesssim 10^6$ K (see the right panel of Fig.\
\ref{fig:Myers78_DM72}). If we then consider a small isochoric (i.e., at
constant density) increase in $T$ of this fluid parcel, we see that
$\Lambda$ decreases. Because we had started from thermal equilibrium, a
drop in $\Lambda$ implies that now $\Gamma > n \Lambda$, a condition
that increases the parcel's temperature even further, causing a runaway
to higher temperatures, until the fluid parcel exits the temperature
range where $(\partial \Lambda / \partial T)_\rho< 0$, at $T \gtrsim
10^6$ K. This behavior is known as {\it thermal instability} (TI), and
the condition $(\partial \Lambda / \partial T)_\rho< 0$ is known as the
{\it isochoric criterion} for TI (Field 1965; see also the review by
V\'azquez-Semadeni et al.\ 2003).

\begin{figure}
\begin{center}
\includegraphics[scale=.7]{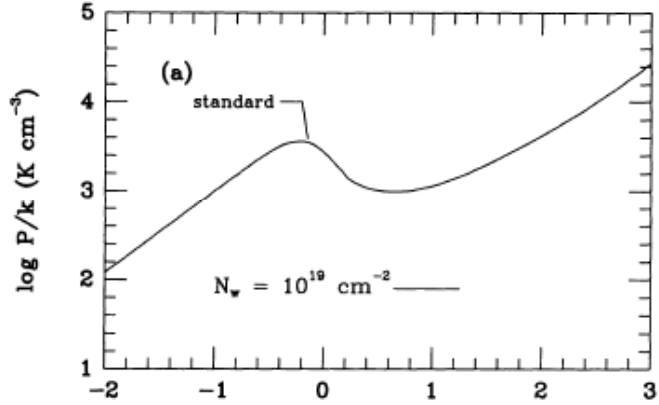}
\end{center}
\caption{Thermal-equilibrium pressure $\Peq$ as a function of number
density for ``standard'' conditions of metallicity and background UV
radiation for the atomic medium. The horizontal axis gives
$\log_{10}(n/{\rm cm}^3)$. From \citet{Wolfire+95}.}
\label{fig:Peq_vs_n}
\end{figure}

Another, less stringent condition for the development of TI is the
so-called {\it isobaric criterion}. This can be most easily understood
as follows. Note that eq.\ (\ref{eq:therm_eq}) provides a relation
between the density and temperature of the medium. This can be inserted
in the equation of state for the gas,\footnote{Given the low densities
of the ISM, it is a good approximation to use the ideal gas equation of
state.} allowing the elimination of the temperature from it and writing a {\it
barotropic} equation of the form $\Peq = \Peq(\rho)$, where $\Peq$ is
the thermal pressure under conditions of thermal equilibrium, and is a
function of the density only. Figure \ref{fig:Peq_vs_n} shows the
resulting density dependence of $\Peq$ for the atomic medium under
``standard'' conditions \citep{Wolfire+95}. The region above the graph
of $\Peq(n)$ corresponds to $n\Lambda > \Gamma$, and therefore to net
cooling. Conversely, the area under the curve corresponds to net
heating, $n \Lambda < \Gamma$.
From this figure, we observe that the slope of the graph of $\Peq$ vs.\
$\rho$ is {\it negative} for densities in the range $0.6 \lesssim n
\lesssim 5 \pcc$ (i.e., $-0.2 \lesssim \log_{10}(n/\pcc) \lesssim
0.7$).

Let us now consider a fluid parcel in this density range, and apply to
it a small, quasi-static compression (i.e., volume reduction) to it,
increasing its density. This displaces the parcel to the region above
the thermal equilibrium curve, where net cooling occurs. Because the
compression is quasi-static, the parcel has time to cool under the
effect of the net cooling, which causes its pressure to decrease as it
attempts to return to the thermal equilibrium curve. But thus it is now
at a lower pressure than its surroundings, which then continue to
further compress the parcel, and runaway compression sets in, until the
parcel exits the regime where $d \Peq / d n < 0$. The density range
where this condition is met is called the {\it unstable range}. 

Thus, if the mean density of the medium is in the unstable range, this
mode of TI tends to cause the medium to spontaneously segregate into a
cold, dense phase and a warm, diffuse one \citep{FGH69}. That is indeed
the case of the ISM in the Galactic midplane in the Solar neighborhood,
a fact which led \citet{FGH69} to propose the so-called {\it two-phase
model} of the ISM: assuming dynamical and thermal equilibrium, the
atomic ISM would consist of small dense, cold clumps (the
CNM), immersed in a warm, diffuse background (the WNM). The clumps are
expected to be small because the fastest growing mode of the instability
occurs at vanishingly small scales in the absence of thermal
conductivity, or at scales $\sim 0.1$ pc for the estimated thermal
conductivity of the ISM \citep[see, e.g.,][]{AH05}. Note that the
isochoric criterion for TI implies that the isobaric one is satisfied,
but not the other way around. Technical and mathematical details, as
well as other modes of TI, can be found in the original paper by
\citet{Field65}, and in the reviews by \citet{Meerson96} and by
\citet{VS+03}. 

It is important to note that, even if the medium is {\it not} thermally
unstable, the balance between heating and cooling implies a certain
functional dependence of $\Peq(\rho)$, which is often approximated by a
{\it polytropic} law of the form $\Peq \propto \rho^\gamef$
\citep[e.g.,][]{Elm91, VPP96}, where it should be noted that $\gamef$,
which we refer to as the {\it effective polytropic exponent}, is {\it not} the
ratio of specific heats for the gas in this case, but rather a free
parameter that depends on the functional forms of $\Lambda$ and
$\Gamma$. The isobaric mode of TI corresponds to $\gamef < 0$.

\section{Compressible Turbulence} \label{sec:turb}

\subsection{Equations} \label{sec:eqs}

In the previous section we have discussed thermal aspects of the ISM,
whose main dynamical effect is the segregation of the medium into the
cold and warm phases. Let us now discuss dynamics. As was mentioned in
Sec.\ \ref{sec:intro}, the ISM is in general highly turbulent, and
therefore it is necessary to understand the interplay between turbulence
and the effects of the net cooling ($n \Lambda - \Gamma$), which affects
the compressibility of the gas \citep{VPP96}.

The dynamics of the ISM are governed by the fluid equations, which,
neglecting magnetic fields, comprise eq.\ (\ref{eq:int_en_cons})
and \citep[e.g.,][]{LL59, Shu92}
\barr
\frac{\partial\rho}{\partial t} + \uu \cdot \nabla \rho &=& -\rho \nabla
\cdot \uu, \label{eq:contin}\\
\frac{\partial\uu}{\partial t} +  \uu\cdot\nabla\uu &=&
-\frac{\nabla P}{\rho} -  \nabla \varphi + \nu \left[\nabla^2 \uu +
\frac{\nabla (\nabla \cdot \uu)}{3} \right],
\label{eq:mom_cons}\\ 
%
%
\nabla^2 \varphi &=& 4 \pi G \rho, \label{eq:Poisson}
\earr
where $\varphi$ is the gravitational potential, and $\nu$ is the
kinematic viscosity. Equation (\ref{eq:contin}) represents mass
conservation, and is also known as the {\it continuity
equation}. Equation (\ref{eq:mom_cons}) is the momentum conservation, or
{\it Navier-Stokes} equation per unit mass, with an additional source
term representing the gravitational force $\nabla \varphi/\rho$. In
turn, the gravitational potential is given by {\it Poisson's equation},
eq.\ (\ref{eq:Poisson}). Equations (\ref{eq:int_en_cons}),
(\ref{eq:contin}), (\ref{eq:mom_cons}), and (\ref{eq:Poisson}) are to be
solved simultaneously, given some initial and boundary conditions.

A brief discussion of the various terms in eq.\ (\ref{eq:mom_cons}) is
in order. The second term on the left is known as the {\it advective}
term, and represents the transport of $i$-momentum by the $j$ component
of the velocity, where $i$ and $j$ represent any two components of the
velocity. It is responsible for {\it mixing}. The pressure gradient term
(first term on the right-hand side [RHS]) in general acts to counteract
pressure, and therefore density, gradients across the flow. Finally, the
term in the brackets on the RHS, the {\it viscous} term, being of a diffusive
nature, tends to erase velocity gradients, thus tending to produce a
uniform flow.

\subsection{Reynolds and Mach numbers. Compressibility} \label{sec:Re_Ms}

Turbulence develops in a flow when the ratio of the advective term to
the viscous term becomes very large. That is,
\beq
\frac{{\cal O}\left[\uu\cdot\nabla\uu \right]} {{\cal O} \left[\nu
\left(\nabla^2 \uu + \frac{\nabla (\nabla \cdot \uu)}{3} \right) \right]} \sim
\frac{U^2}{L} \left[\nu \frac{U}{L^2}\right]^{-1} \sim \frac{UL}{\nu}
\equiv \Rey \gg 1,
\label{eq:Reynolds_num}
\eeq
where $\Rey$ is the {\it Reynolds number}, $U$ and $L$ are
characteristic velocity and length scales for the flow, and ${\cal O}$
denotes ``order of magnitude''. This condition
implies that the mixing action of the advective term overwhelms the
velocity-smoothing action of the viscous term.

On the other hand, noting that the advective and pressure gradient
terms contribute comparably to the production of density fluctuations, we
can write
\barr
1 \sim \frac{{\cal O}\left(\uu\cdot\nabla\uu \right)} {{\cal O}\left(\nabla P/\rho \right)} \sim \frac{U^2}{L}
\left[\frac{\Delta P}{L \rho} \right]^{-1} &\sim& U^2 \left(\frac{\cs^2
\Delta \rho}{\rho} \right)^{-1} \equiv \Ms^2 \left(\frac{\Delta \rho}
{\rho} \right)^{-1},\\
\Rightarrow \frac{\Delta \rho}{\rho} &\sim& \Ms^2 \label{eq:dens_jumps},
\earr
where $\Ms \equiv U/\cs$ is the {\it sonic Mach number}, and we have
made the approximation 
that $\Delta P/\Delta \rho \sim \cs^2$, where $\cs$ is the sound
speed. Equation (\ref{eq:dens_jumps}) then implies that strong
compressibility requires $\Ms \gg 1$. Conversely, flows with $\Ms \ll 1$
behave incompressibly, even if they are gaseous. Such is the case, for
example, of the Earth's atmosphere. In the incompressible limit, $\rho
=$ cst., and thus eq.\ (\ref{eq:contin}) reduces to $\nabla \cdot \uu =
0$.

Finally, a trivial, but often overlooked, fact is that, in order to
produce a density enhancement in a certain region of the flow, the
velocity at that point must have a negative divergence (i.e., a {\it
convergence}), as indicated by the continuity equation, eq.\
(\ref{eq:contin}). It is very frequent to encounter in the literature
discussions of pre-existing density enhancements (``clumps'') in
hydrostatic equilibrium. But it should be kept in mind that these can
only exist in multi-phase media, where a dilute, warm phase can have the
same pressure as a denser, but colder, clump. But even in this case, the
{\it formation} of that clump must have initially involved the
convergence of the flow towards the cloud, and the hydrostatic situation
is applicable in the limit of very long times after the formation of the
clump, when the convergence of the flow has subsided.

\subsection{Production of Density Fluctuations} \label{sec:dens_fluc}

According to the previous discussion, a turbulent flow in which the
velocity fluctuations are supersonic will naturally develop strong
density fluctuations.  Note, however,
that the nature of turbulent density fluctuations in a single-phase
medium (such as, for example, a regular isothermal or adiabatic flow)
is very different from that of the cloudlets formed by TI (cf.\ Sec.\
\ref{sec:thermodynamics}). In a single-phase turbulent medium, turbulent
density fluctuations must be transient, as a higher density generally
conveys a higher pressure,\footnote{An exception would be a so-called
Burgers' flow, which is characterized by the absence of the pressure
gradient term \citep{Burgers74}.} and therefore the fluctuations must
re-expand after the compression that produced them has subsided.

For astrophysical purposes it is important to determine the distribution
of these fluctuations, as they may constitute, or at least provide the
seeds for, what we normally refer to as ``clouds'' in the ISM. Because of
the transient nature of turbulent density fluctuations in single-phase
media, however, this distribution refers to a time-stationary population
of fluctuations, although the fluctuations themselves will appear and
disappear over times short compared to the time over which the
distribution is considered.

The probability density distribution (PDF) of the density field in
turbulent isothermal flows was first investigated numerically, finding
that, in the isothermal case, it posesses a lognormal form \citep{VS94}.
A theory for the emergence of this functional form was later proposed by
\citet{PV98}, in which the production of density fluctuations was
assumed to arise from a succession of compressive or expansive waves,
each one acting on the value of the density left by the previous one.
Because the medium contains a unique distribution of (compressible)
velocity fluctuations, and because the density jumps in isothermal flow
depend only on Mach number but not in the local density, the density
fluctuations belong all to a unique distribution as well, yet each one
can be considered independent of the others if the global time scales
considered are much longer than the autocorrelation time of the velocity
divergence \citep{Blaisdell+93}. Finally, because the density jumps are
multiplicative in the density (cf.\ eq.\ \ref{eq:dens_jumps}), then they
are additive in $s \equiv \ln \rho$. Under these conditions, the Central
Limit Theorem can be invoked for the increments in $s$, implying that
$s$ will be normally distributed. In consequence, $\rho$ will have a
lognormal PDF. 

\begin{figure}
\begin{center}
\includegraphics[scale=.4]{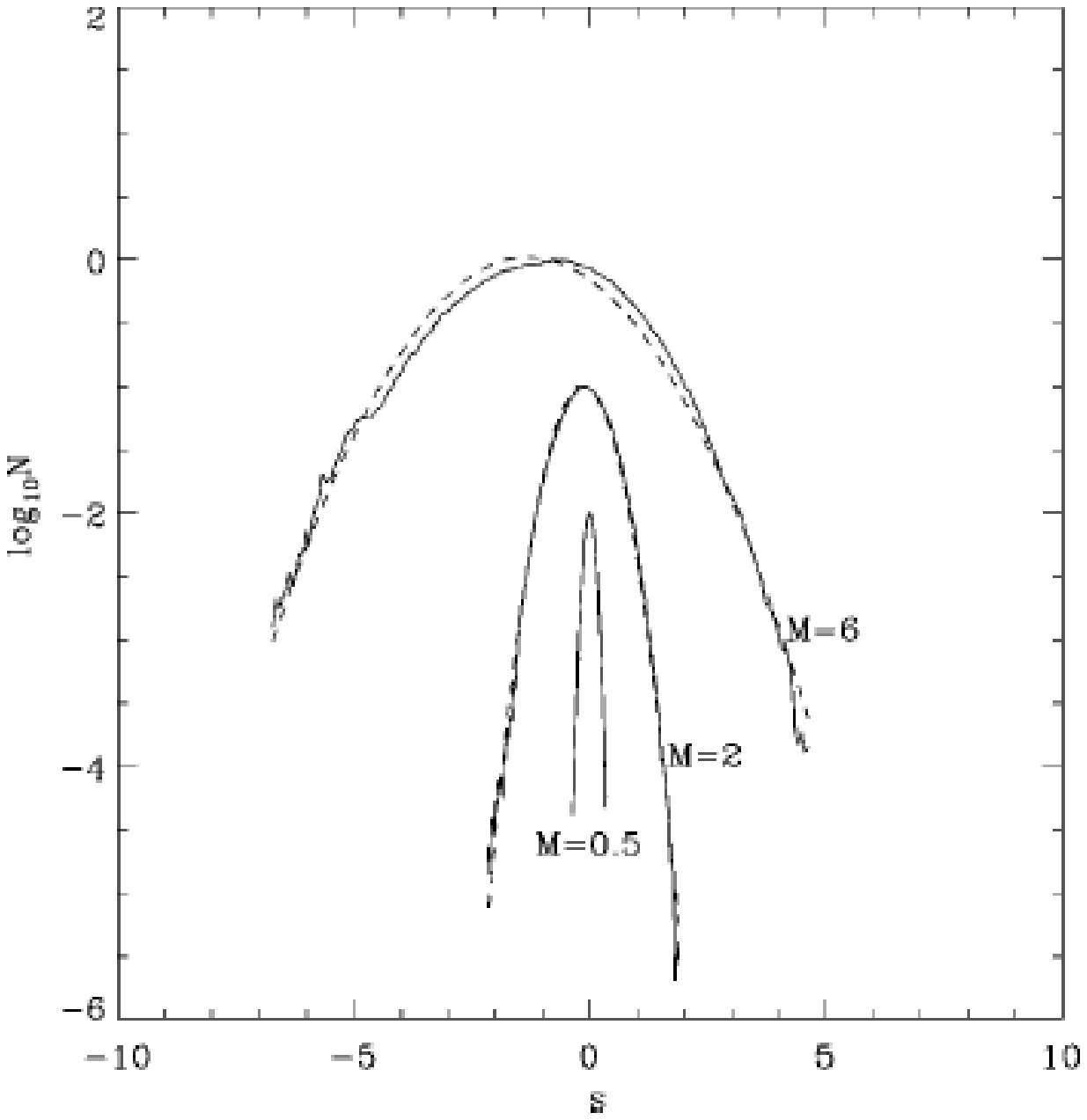}
\includegraphics[scale=.29]{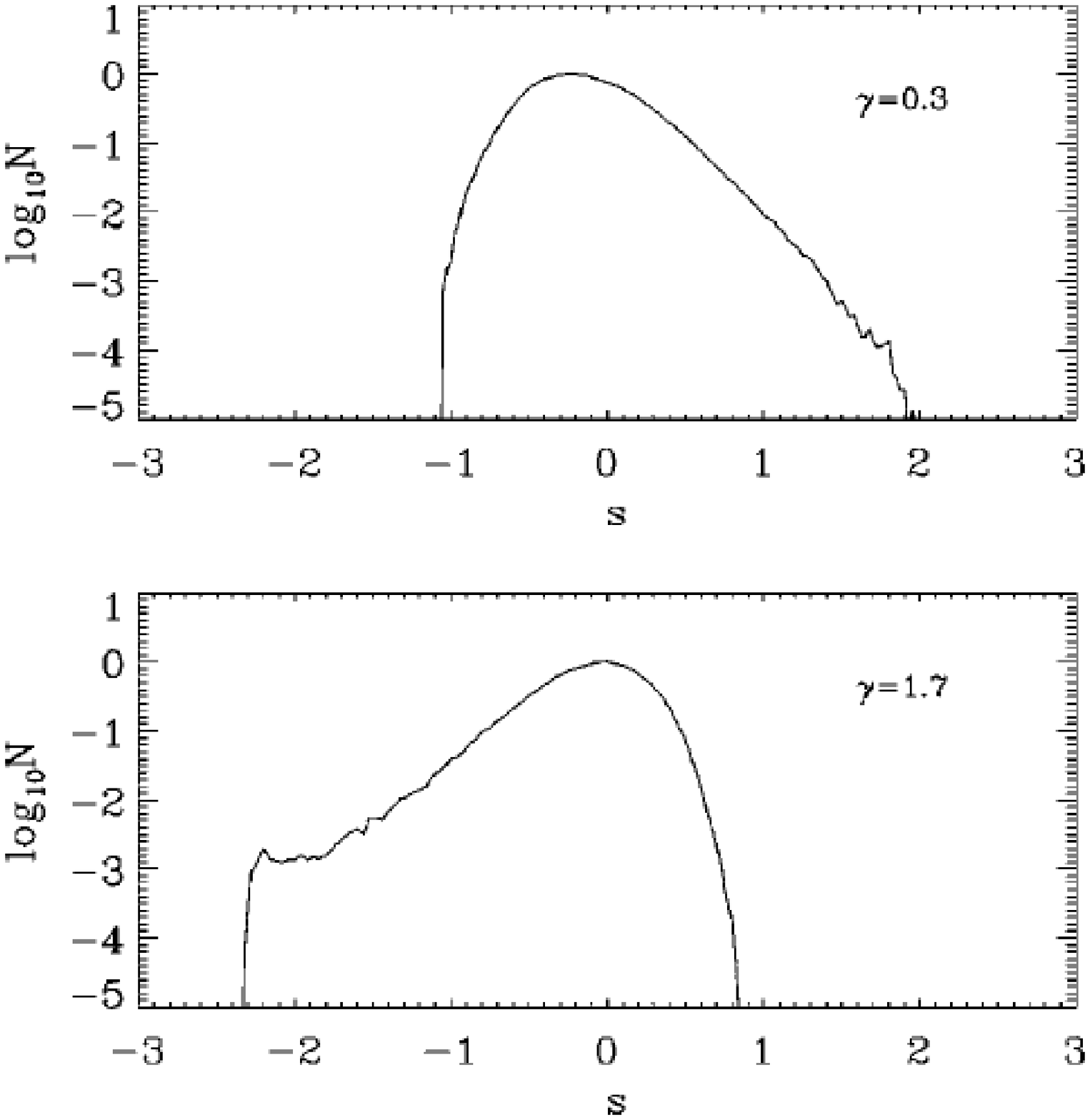}
\end{center}
\caption{{\it Left:} Lognormal density PDFs for isothermal
one-dimensional simulations at various Mach numbers, indicated by the
labels. The independent variable is $s \equiv \ln \rho$. {\it Right:}
Density PDFs for polytropic cases, with effective polytropic exponent
$\gamef = 0.3$ ({\it top}) and $\gamef = 1.7$ ({\it bottom}). From
\citet{PV98}.}
\label{fig:pdfs_PV98}
\end{figure}

In addition, \citet{PV98} also argued that the variance of the density
fluctuations should scale linearly with $\Ms$, a fact that has been
investigated further by various groups \citep{Padoan+97,
PV98, Federrath+08}. In particular, using numerical simulations of
compressible turbulence driven by either solenoidal (or ``vortical'') or
compressible (or ``potential'') forces, the latter authors proposed that the
variance of $s$ is given by 
\beq 
\sigma_s = \left[\ln(1 + b \Ms^2)\right],
\label{eq:sigma_Ms}
\eeq
where $b$ is a constant whose value depends on the nature of the forcing,
taking the extreme values of $b=1/3$ for purely solenoidal forcing, and
$b=1$ for purely compressible forcing. The lognormal density PDF for
the one-dimensional simulations of \citet{PV98}, with
its dependence on $\Ms$, is illustrated in the {\it left panel} of Fig.\
\ref{fig:pdfs_PV98}. 

Finally, \citet[][see also Padoan \& Nordlund 1999]{PV98} also
investigated the case where the flow behaves as a polytrope with
arbitrary values of $\gamef$, by noting that in this case the sound
speed is not constant, but rather depends on the density as $\cs \propto
\rho^{(\gamef -1)/2}$, implying that the local Mach number of a fluid
parcel now depends on the local density besides its dependence on the
value of the flow velocity. Introducing this dependence of $\Ms$ on
$\rho$ in the expression for the lognormal PDF, \citet{PV98} concluded
that the density PDF should develop a power-law tail, at high densities
when $\gamef <1$, and at low densities when $\gamef > 1$. This result
was then confirmed by numerical simulations of polytropic turbulent
flows (Fig.\ \ref{fig:pdfs_PV98}, {\it right panel}).

\section{Turbulence and Thermodynamics} \label{sec:turb_therm}

In the previous sections we have separately discussed two different
kinds of physical processes operating in the ISM: radiative heating and
cooling (to which we refer collectively as {\it net cooling}, $n \Lambda -
\Gamma$), and compressible turbulence. However, since both operate
simultaneously, it is important to understand how they interact with
each other, especially because the mean density in the Solar
neighborhood, $\langle n \rangle \sim 1 \pcc$, falls precisely in the
thermally unstable range. This problem has been investigated numerically
by various groups \citep[e.g.,][]{HP99, WF00, KI00, KI02, VS+00a, VS+03,
VS+06, Gazol+01, Gazol+05, KN02, SS+02, PO04, PO05, AH05, Heitsch+05, HA07}.

\subsection{Density and Pressure Distributions in the ISM}
\label{sec:pres_distr} 

\begin{figure}
\begin{center}
\includegraphics[scale=.27]{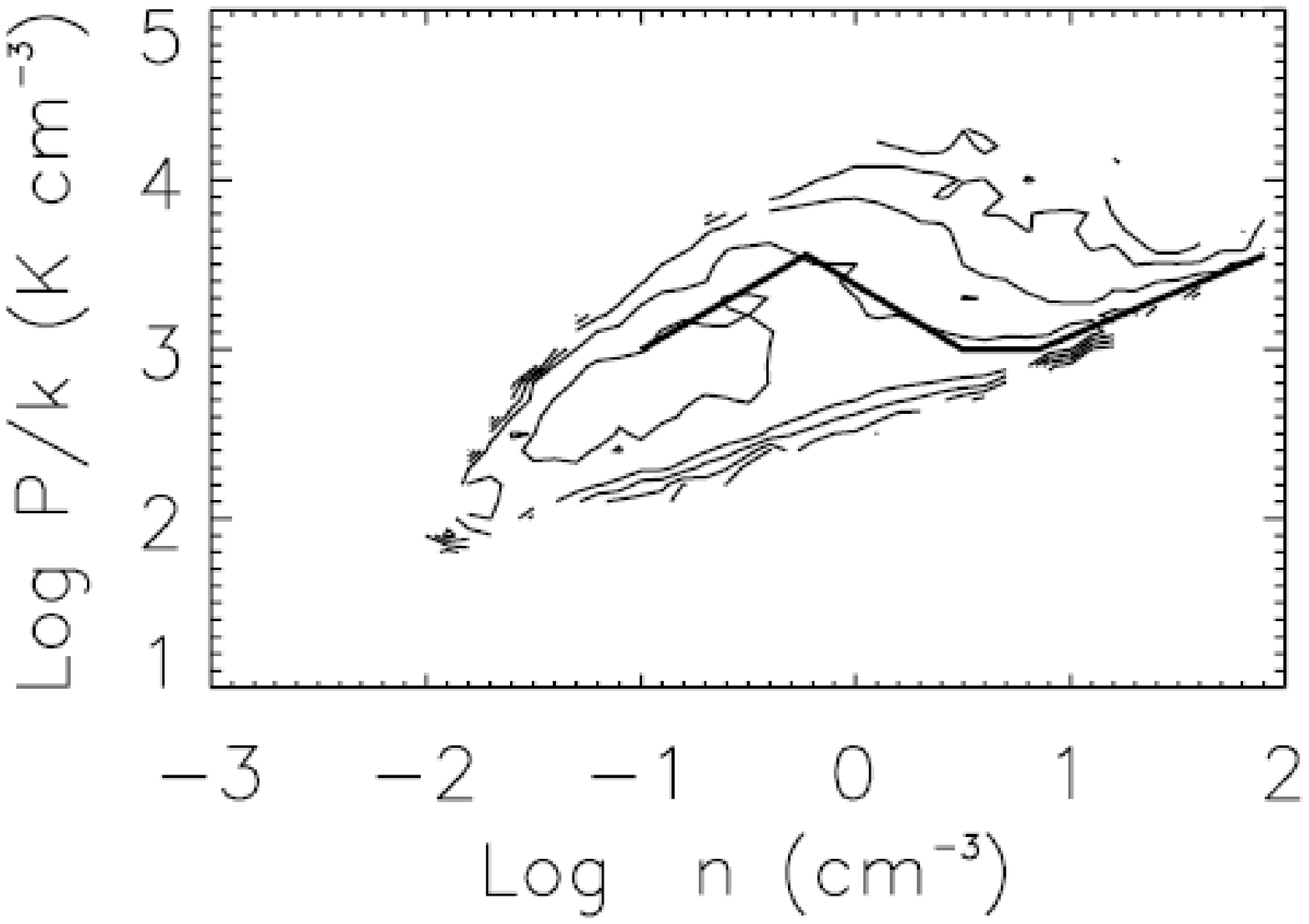}
\includegraphics[scale=.27]{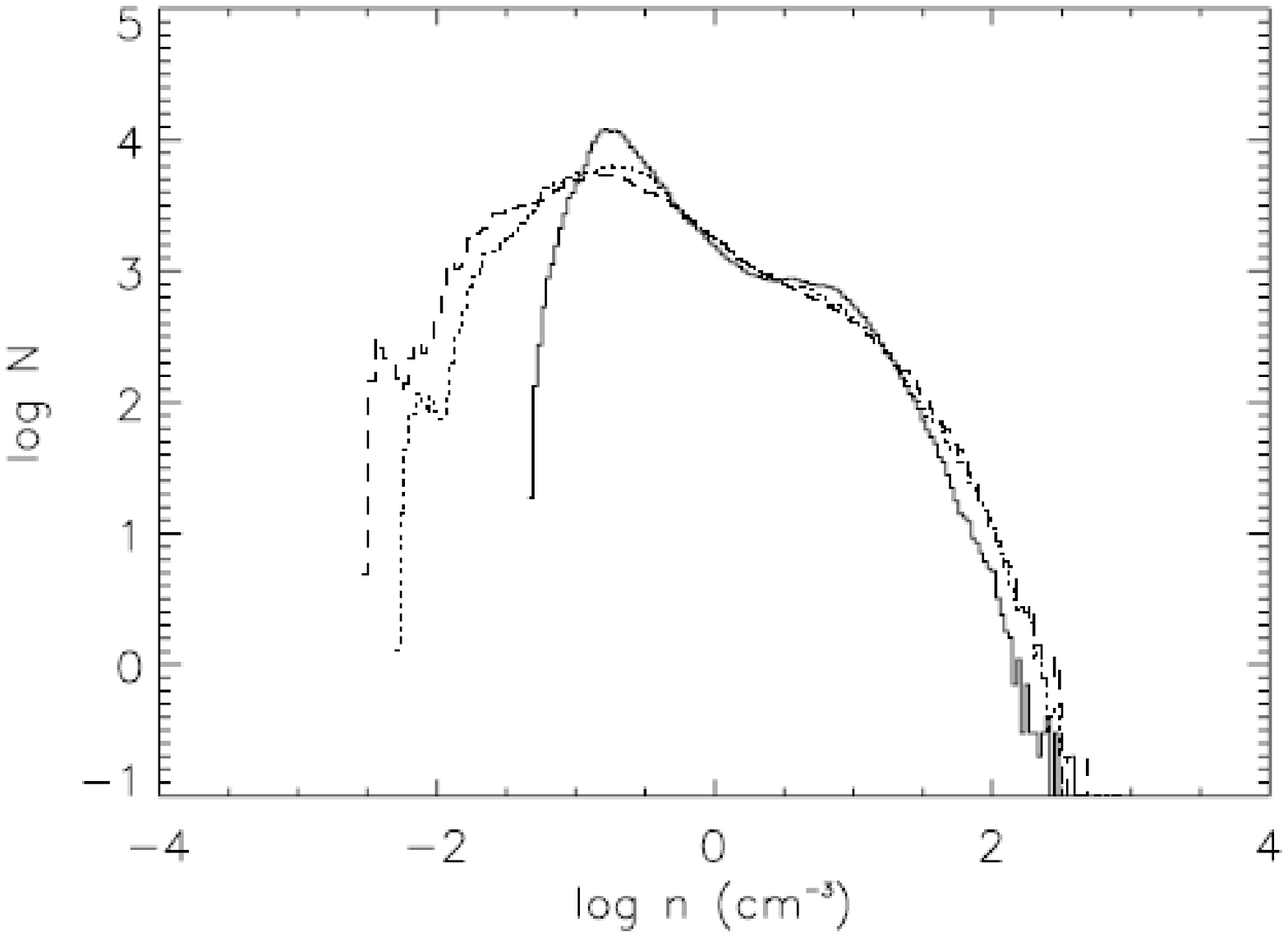}
\end{center}
\caption{{\it Left:} Two-dimensional histogram of the grid cells in the
pressure-density diagram for a two-dimensional simulation of turbulence
in the thermally-bistable atomic medium, with {\it rms} velocity
dispersion of $9 \kms$, a numerical box size of 100 pc, and the
turbulent driving applied at a scale of 50 pc. {\it Right:} Density PDF
in simulations like the one on the left panel, but with three different
values of the {\it rms} velocity: $4.5 \kms$ ({\it solid line}), $9
\kms$ ({\it dotted line}), and $11.3 \kms$ ({\it dashed line}). From
\citet{Gazol+05}.
}
\label{fig:P_rho_pdfs_TI}
\end{figure}

The main parameter controlling the interaction between turbulence and
net cooling is the ratio $\eta \equiv \tcool/\tturb$, where $\tcool
\approx e/(\mu \mh \rho \Lambda)$ is the cooling time and $\tturb
\approx L/U$ is the turbulent crossing time. The remaining symbols have
been defined above. In the limit $\eta \gg 1$, the turbulent
compressions evolve much more rapidly than they can cool, and therefore
behave nearly adiabatically. Conversely, in the limit $\eta \ll 1$, the
fluctuations cool down essentially instantaneously while the turbulent
compression is evolving, and thus they tend to reach the thermal
equilibrium pressure $\Peq$ as soon as they are produced\footnote{Note
that it is often believed that fast cooling implies isothermality.
However, this is an erroneous notion. While it is true that fast cooling
is a necessary condition for isothermal behavior, the reverse
implication does not hold. Fast cooling only implies an approach to the
thermal equilibrium condition, but this need not be
isothermal. The precise form of the effective equation of state depends
on the details of the functional dependence of $\Lambda$ and $\Gamma$ on
$T$ and $\rho$.} \citep{Elm91, PVP95, SS+02, VS+03, Gazol+05}. Because in a
turbulent flow velocity fluctuations of a wide range of amplitudes and
size scales are present, the resulting density fluctuations in general
span the whole range between those limits, and the actual thermal
pressure of a fluid parcel is not uniquely determined by its density,
but rather depends on the details of the velocity fluctuation that
produced it. This causes a scatter in the values of the pressure around
the thermal-equilibrium value in the pressure-density diagram (Fig.\
\ref{fig:P_rho_pdfs_TI}, {\it left panel}), and also produces
significant amounts of gas (up to nearly half of the total mass) with
densities and temperatures in the classically forbidden thermally
unstable range \citep{Gazol+01, dAB05, AH05, ML+05}, a result that has
been encountered by various observational studies
\citep[e.g.,][]{Dickey+78, Heiles01}. In any case,  the
tendency of the gas to settle in the stable 
phases still shows up as a multimodality of the density PDF, which
becomes less pronounced as the {\it rms} turbulent velocity increases
(Fig.\ \ref{fig:P_rho_pdfs_TI}, {\it right panel}).

\subsection{The formation of dense, cold clouds} \label{sec:cloud_form}

\begin{figure}
\begin{center}
\includegraphics[scale=.3]{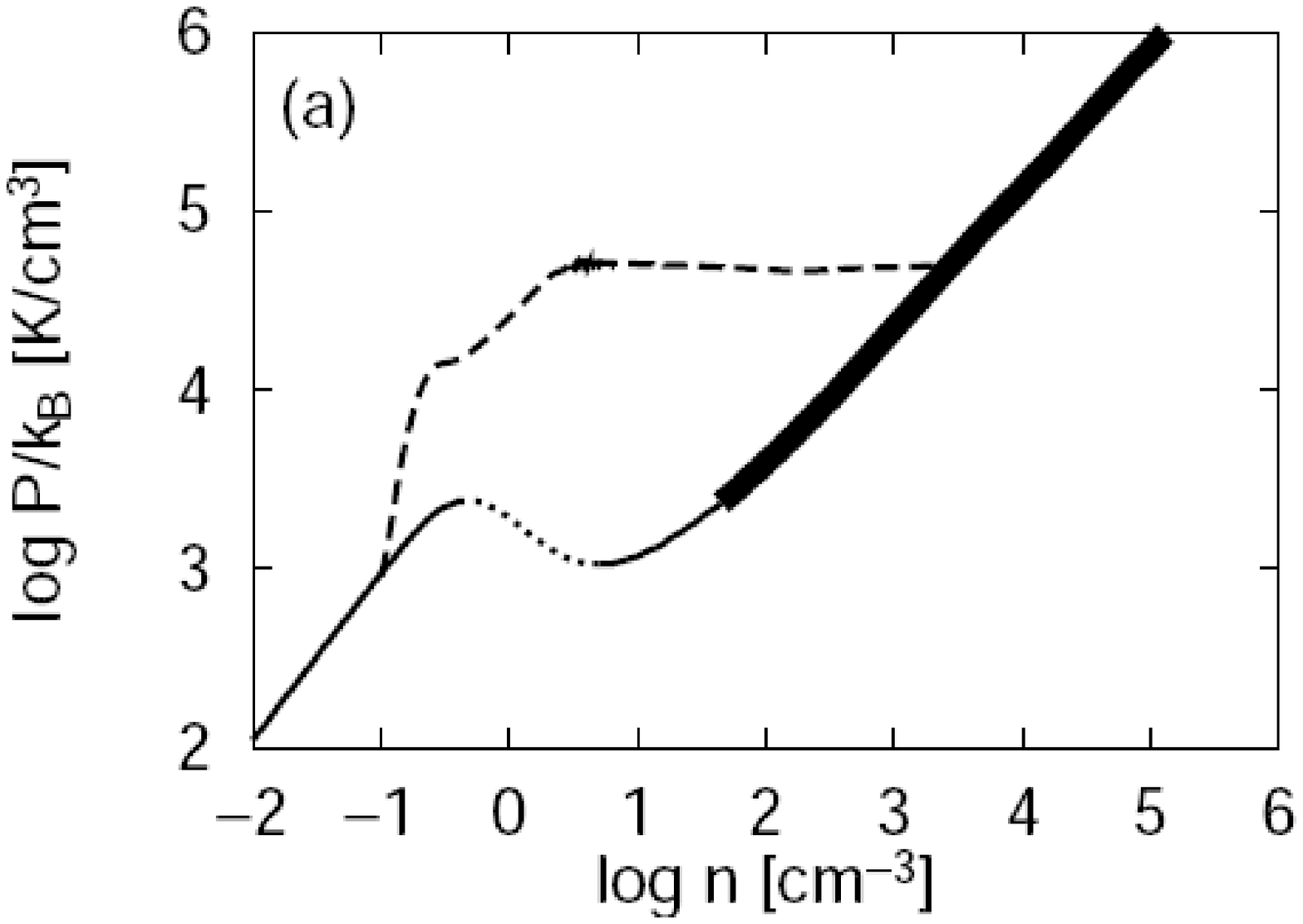}
\includegraphics[scale=.27]{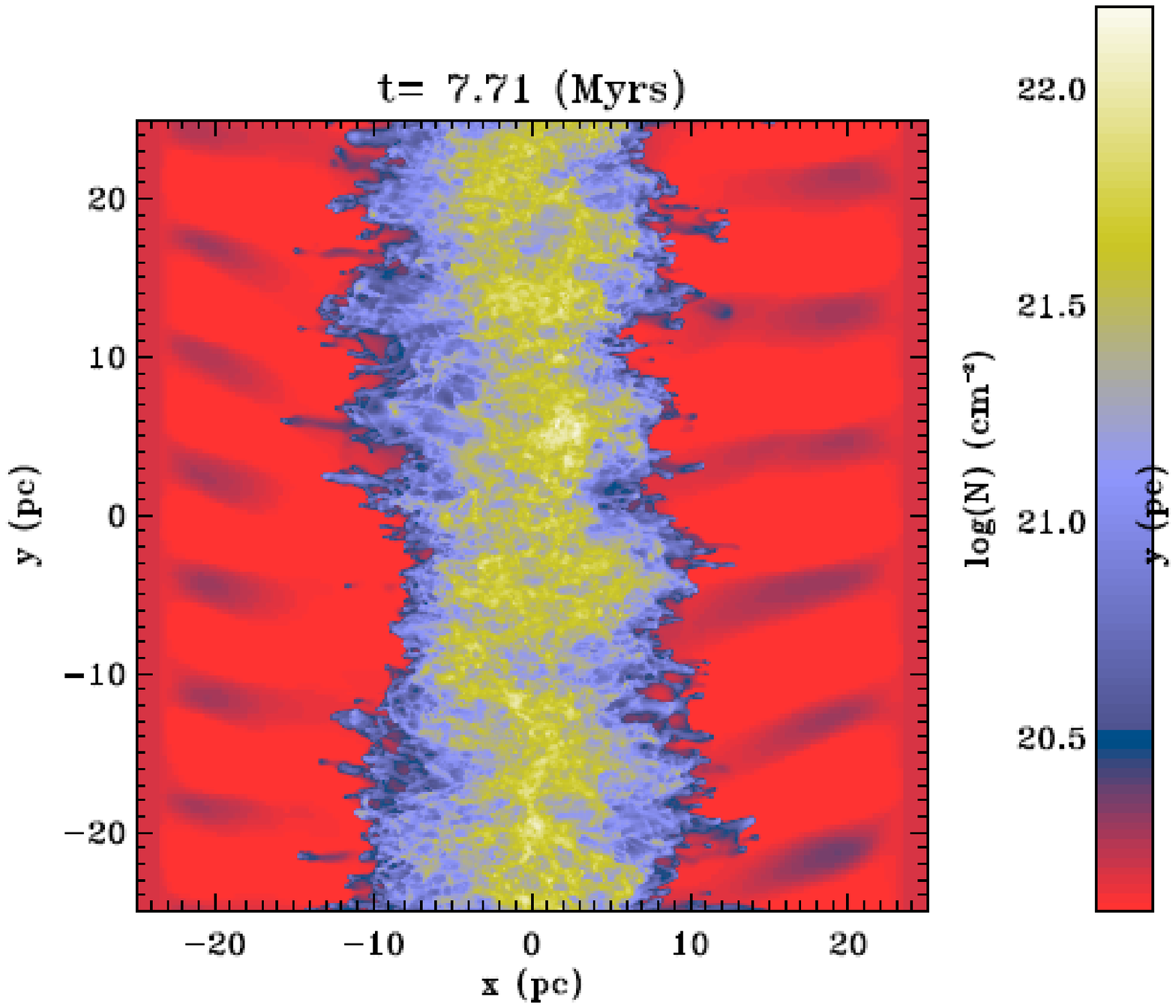}
\end{center}
\caption{{\it Left:} Evolutionary path ({\it dashed line}) in the $P$
vs.\ $\rho$ diagram of a fluid parcel initially in the WNM after
suffering a transonic compression that nonlinearly triggers TI. The {\it
solid} and {\it dotted lines} show the locus of $\Peq(\rho)$, the solid
sections corresponding to linear stability and the dotted ones to linear
instability. The solid section to the left of the dotted line
corresponds to the WNM and the one at the right, to the CNM. The
perturbed parcel evolves from left to right along the dashed line. From
\citet{KI00}. {\it Right:} Projected (or {\it column}) density of a
numerical simulation of the formation of a dense cloud formed by the
convergence of two large-scale streams of WNM. The projection is along
lines of sight perpendicular to the direction of compression. The cloud
is seen to have become turbulent and highly fragmented. From
\citet{Hennebelle+08}.} 
\label{fig:nonlin_condens}
\end{figure}

Another important consequence of the interaction of ``turbulence'' (or,
more generally, large-scale coherent motions of any kind) and TI is that
the former may {\it nonlinearly} induce the latter. Indeed, \citet[][see
also Koyama \& Inutsuka 2000]{HP99} showed that transonic (i.e., with
$\Ms \sim 1$) compressions in the WNM can compress the medium and bring it
sufficiently far from thermal equilibrium that it can then undergo a
phase transition to the CNM (Fig.\ \ref{fig:nonlin_condens}, {\it left
panel}). This process amounts then to producing a cloud with a density
up to $100\times$ larger than that of the WNM by means of only moderate
compressions. This is in stark contrast with the process of producing
density fluctuations by pure supersonic compressions in, say, an
isothermal medium, in which such density contrasts would require Mach
numbers $\Ms \sim 10$.

Moreover, the cold clouds formed by this mechanism have typical sizes
given by the size scale of the compressive wave in the transverse
direction to the compression, thus avoiding the restriction of having
the size scale of the fastest growing mode of TI, which is very small
($\sim 0.1$ pc; cf.\ Sec.\ \ref{sec:thermodynamics}). The initial stages
of this process may produce thin CNM sheets \citep{VS+06}, which are in
fact observed \citep{HT03}. However, such sheets are quickly
destabilized, apparently by a combination of nonlinear thin shell
\citep[NTSI;][]{Vishniac94}, Kelvin-Helmholtz and Rayleigh-Taylor
instabilities \citep{Heitsch+05}, fragmenting and becoming
turbulent. This causes the clouds to become a complex mixture of cold
and warm gas, where the cold gas is distributed in an intrincate network
of sheets, filaments and clumps, possibly permeated by a dilute, warm
background. An example of this kind of structure is shown in the {\it
right panel} of Fig.\ \ref{fig:nonlin_condens}.

\section{Turbulence in the Ionized ISM} \label{sec:turb_diff}

As discussed in the previous sections, the ionized and atomic components
of the ISM consist of gas in a wide range of temperatures, from $T \sim
10^6$ K for the HIM, to $T \sim 40$ K for the CNM. In particular,
\citet{HT03} report temperatures in the range $500 < T < 10^4$ K for the
WNM, and in the range $10 < T < 200 $ K for the CNM. The WIM is expected
to have $T \sim 10^4$ K. Additionally, those same authors report column
density-weighted {\it rms} velocity dispersions $\sigma_v \sim 11 \kms$
for the WNM, and $\sigma_v \sim 7 \kms$ for the CNM. Since the adiabatic
sound speed is given by \citep[e.g.,][]{LL59}
\beq 
\cs = \sqrt{\frac{\gamma k T}{\mu \mh}} =  10.4 \kms \left(\frac{T}
{10^4 {\rm K}} \right)^{1/2},
\label{eq:cs}
\eeq
it is clear that the warm, or {\it diffuse}, gas is transonic ($\Ms \sim
1$), while the cold, or {\it dense}, gas is strongly supersonic ($3
\lesssim \Ms \lesssim 20$). Indeed, collecting measurements of
interstellar scintillation (fluctuations in amplitude and phase of radio
waves caused by scattering in the ionized ISM) from a variety of
observations, \citet{Armstrong+95} estimated the power spectrum of
density fluctuations in the WIM, finding that it is consistent with a
Kolmogorov spectrum, characteristic of incompressible turbulence
\citep[see, e.g., the reviews by][]{VS99, MK04, ES04}, on
scales $10^8 \lesssim L \lesssim 10^{15}$ cm. 

More recently, using data
from the Wisconsin H$\alpha$ Mapper Observatory, \citet{CL10} have been
able to extend this result to scales $\sim 10^{19}$ cm, suggesting that
the WIM behaves as an incompressible turbulent flow over size scales
spanning more than 10 orders of magnitude. Although the WIM is ionized,
and thus should be strongly coupled to the magnetic field, the
turbulence then being magnetohydrodynamic (MHD), Kolmogorov scaling
should still apply, according to the theory of incompressible MHD
fluctuations \citep{GS95}. The likely sources of kinetic energy for
these turbulent motions are stellar energy sources such as supernova
explosions \citep[see, e.g.,][]{MK04}.

\section{``Turbulence'' in the Atomic and Molecular ISM} \label{sec:turb_dense}

\subsection{The atomic medium} \label{sec:atomic_turb}

In contrast to the relatively clear-cut situation for the ionized ISM,
the turbulence in the neutral (atomic and molecular) gas is more
complicated, and is currently under strong debate. According to the
discussion in Sec.\ \ref{sec:turb_diff}, the temperatures in the atomic
gas may span a continuous range from a few tens to several thousand
degrees, and have velocity dispersions of $\sigma_v \sim 7$--$10\kms$,
suggesting that it should range from mildly to strongly supersonic. But
because the atomic gas is {\it thermally bistable} (i.e., has two stable
thermodynamic phases separated by an unstable one; Sec.\
\ref{sec:thermodynamics}), and because transonic compressions in the WNM
can nonlinearly induce TI and thus a phase transition to the CNM (Sec.\
\ref{sec:pres_distr}), the neutral atomic medium is expected to
consist of a complex mixture of gas spanning over two orders of
magnitude in density. Early models \citep[e.g.,][]{FGH69, MO77} proposed
that the phases were completely separate, but the results reported in
Sec.\ \ref{sec:pres_distr} suggest that significant amounts of gas exist
as well in the unstable range, transiting between the stable
phases. Numerical simulations of such systems suggest that the velocity
disperion {\it within} the densest ``clumps'' is subsonic, but that the
velocity dispersion of the clumps within the diffuse substrate is
supersonic with respect to the clumps' sound speed \citep[although
subsonic with respect to the warmest gas;][]{KI02, Heitsch+05}. Also,
note that, contrary to earlier ideas \citep[e.g.][]{Kwan79, BS80}, the
clumps in the modern simulations actually form from {\it fragmentation}
of large-scale clouds formed by large-scale compressive motions in the WNM,
rather than the clouds forming from random collision and coagulation of
the clumps. The complexity of this type of structure is illustrated in
the {\it right panel} of Fig.\ \ref{fig:nonlin_condens}.

\subsection{The Molecular Gas} \label{sec:molec_turb}

The discussion so far, involving mainly turbulence and thermodynamics,
has referred to the ionized and atomic components of the ISM. However,
molecular clouds (MCs) have long been known to be strongly self-gravitating
\citep[e.g.,][]{GK74, Larson81}. In view of this, \citet{GK74} initially
proposed that MCs should be in a state of gravitational collapse, and
that the observed motions in MCs (as derived by the non-thermal
linewidths of molecular lines) corresponded to this collapse.
However, shortly thereafter, \citet{ZP74} argued against this
possibility by noting that, if all the molecular gas in the Galaxy
($\Mmol \sim 10^9 \Msun$) were in free-fall, a simple estimate of the
Galaxy's star formation rate (SFR), given by SFR $\sim \Mmol / \tff \sim
200~\Msun$ yr$^{-1}$, where $\tff = \sqrt{3\pi/32 G \rho}$ is the
free-fall time, would exceed the observed rate of $\sim 3~\Msun$
yr$^{-1}$ by roughly two orders of magnitude. This prompted the
suggestion \citep{ZE74} that the non-thermal motions in MCs corresponded
instead to small-scale (in comparison to the clouds' sizes) random 
turbulent motions. The need for these motions to be confined to small
scales arose from the need of the turbulent ({\it ram}) pressure to
provide an isotropic pressure that could counteract the clouds'
self-gravity at large, maintaining them in near virial equilibrium
\citep{Larson81}. Because turbulence is known to be a
dissipative phenomenon \citep[e.g., ][]{LL59}, research then focused on
finding suitable sources for driving the turbulence and avoiding rapid
dissipation. The main driving source was considered to be energy
injection from stars \citep[e.g.,][]{NS80, McKee89, MK04}, and reduction
of dissipation was proposed to be accomplished by having the turbulence
being MHD, and consisting mostly of Alfv\'en waves, which do not
dissipate as rapidly \citep[e.g.,][]{SAL87}.

\begin{figure}
\begin{center}
\includegraphics[scale=.45]{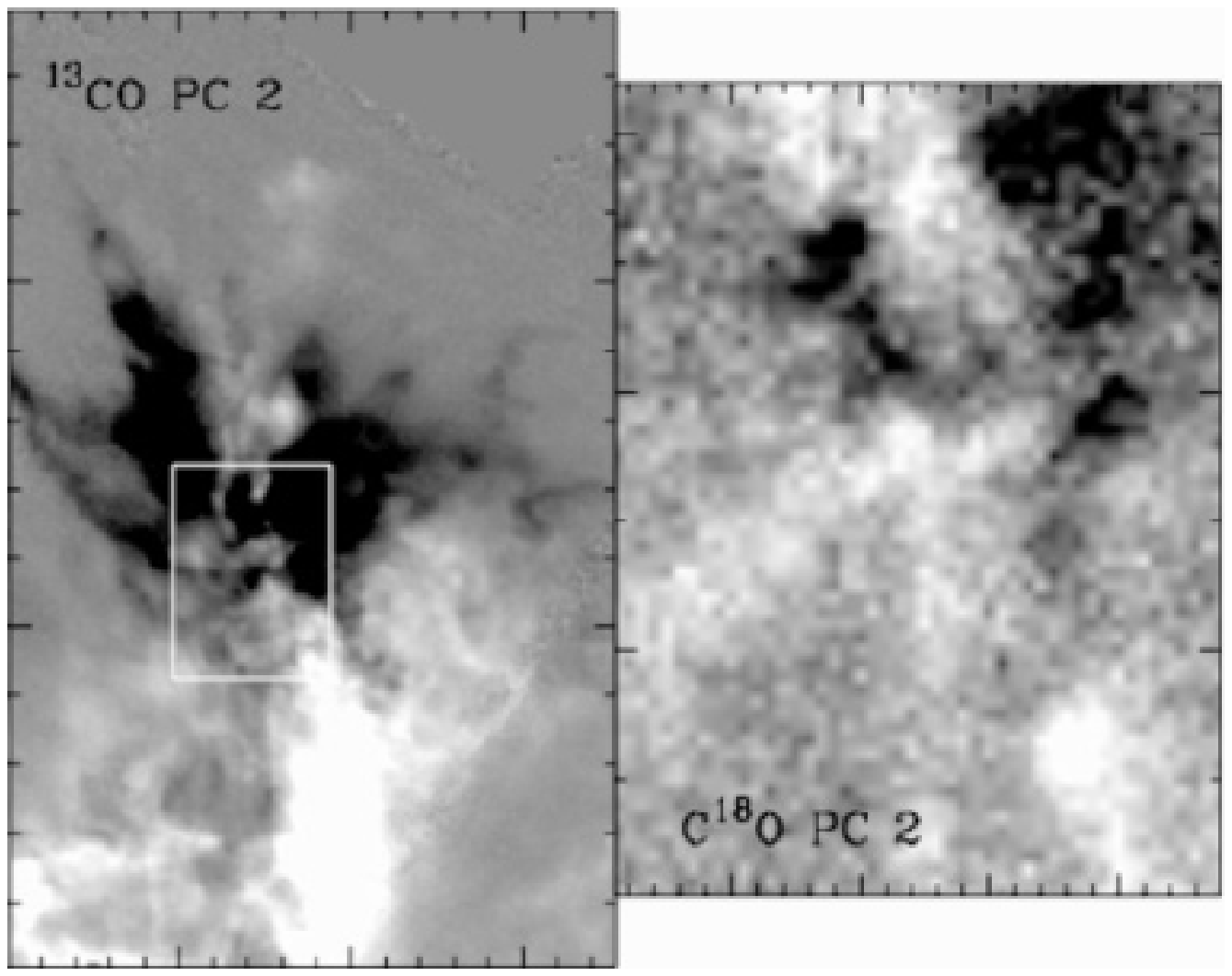}
\includegraphics[scale=.2]{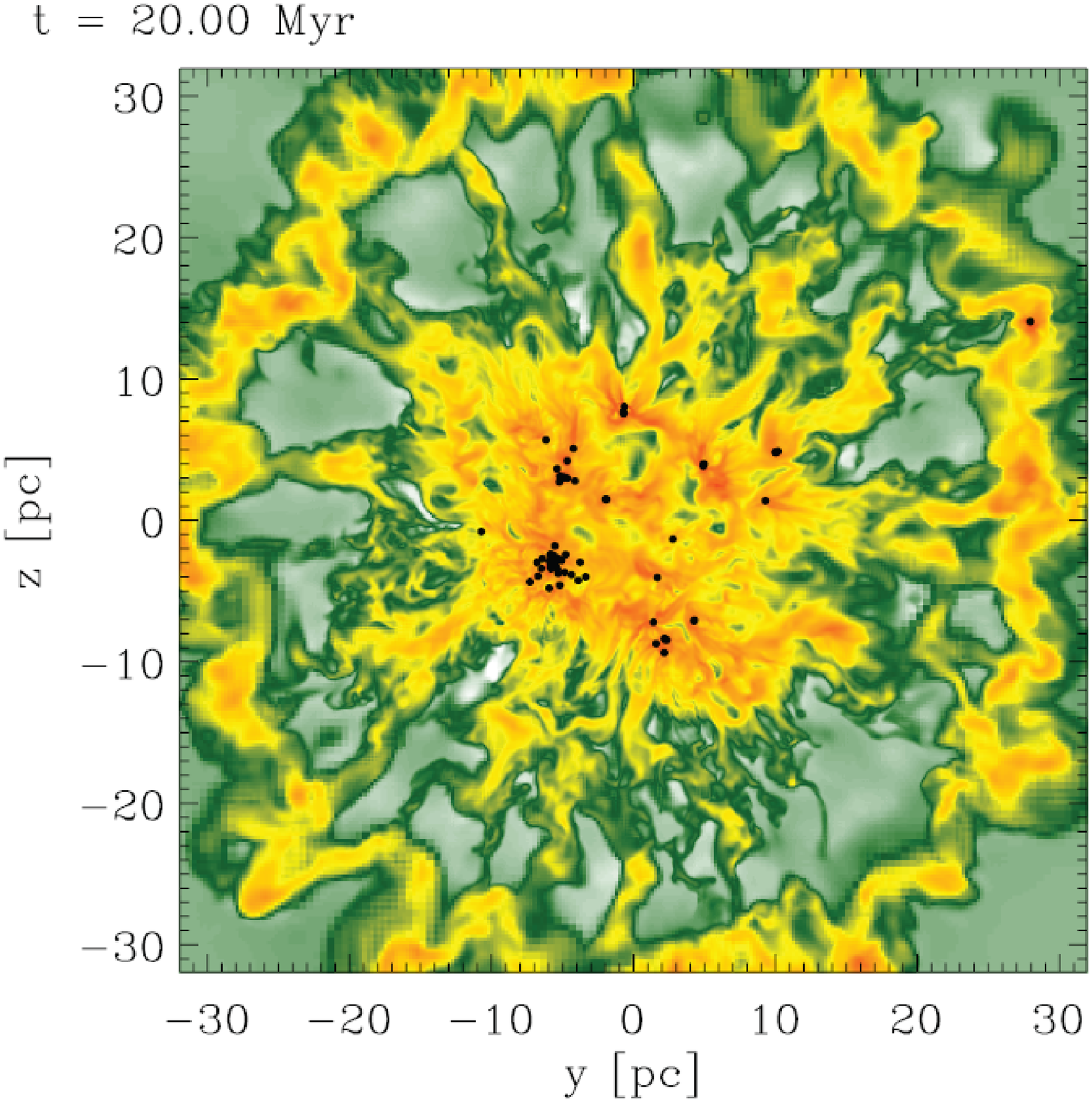}
\end{center} 
\caption{{\it Left and middle panels:} Second eigenimages obtained by
Principal Component Analysis of spectroscopic data of the star-forming
region NGC 1333, showing the main contribution to the linewidth of
molecular emission in this region \citep{Brunt+09}. The middle image
shows the region enclosed in the rectangle in the left image. Black and
white colors represent oppositely-signed components of the
velocity. \citet{Brunt+09} describe the pattern as a ``dipole'', in
which large-scale patches of alternating velocity direction are
observed. This is seen in both the large-scale and the small-scale
images. {\it Right panel:} Image of the projected density field of a 3D
numerical simulation with cooling, self-gravity, and magnetic fields,
representing the formation of a dense atomic cloud by the collision of
WNM streams in the direction perpendicular to the plane of the
figure. The time shown is 20 Myr after the start of the simulation. The
black dots denote ``sink'' particles, which replace local collapsing
zones in the simulation. The whole cloud is also collapsing, although
its collapse is not completed yet by the end of the simulation, at
$t=31$ Myr. From \citet{VS+11}.}
\label{fig:dense_turb}
\end{figure}

However, in the last decade several results have challenged the
turbulent pressure-support scenario: 1) Turbulence is known to be
characterized by having the largest-velocities occur at the largest
scales, and MCs are no exception, exhibiting scaling relations between
velocity dispersion and size suggesting that the largest dispersions
tend to occur at the largest scales \citep[][Fig.\ \ref{fig:dense_turb},
{\it left and middle panels}] {Larson81, HB04, Brunt+09}. This is
inconsistent with the small-scale requirement for turbulent support. 2)
It was shown by several groups that MHD turbulence dissipates just as
rapidly as hydrodynamic turbulence \citep{ML+98, Stone+98, PN99},
dismissing the notion of reduced dissipation in ``Alfv\'en-wave
turbulence'', and thus making the presence of strong driving sources for
the turbulence an absolute necessity. 3) Clouds with very different
contributions from various turbulence-driving mechanisms, including
those with little or no star formation activity, such as the so-called
{\it Maddalena's cloud}, show similar turbulence characteristics
\citep{Williams+94, Schneider+11}, suggesting that stellar energy
injection is not the main source of turbulence in MCs.

Moreover, simulations of dense cloud formation have shown that, once a
large cold CNM cloud forms out of a collision of WNM streams, it quickly
acquires a large enough mass that it can begin to collapse
gravitationally \citep{VS+07, VS+10, VS+11, Heitsch+08,
Heitsch+08b}. The enhancement in its column density promotes the
formation of molecular hydrogen (H$_2$) \citep{HBB01, Bergin+04,
HH08}. Thus, it appears that the formation of a {\it molecular} cloud
may require previous gravitational contraction \citep[see
also][]{McKee89}. In addition, according to the discussion in Secs.\
\ref{sec:cloud_form} and \ref{sec:atomic_turb}, the CNM clouds formed by
converging WNM flows should be born turbulent and clumpy. The turbulent
nature of the clouds further promotes the formation of molecular
hydrogen \citep{GM07}. The simulations by \citep{VS+07, VS+10, VS+11}
show that the nonlinear, turbulent density fluctuations can begin to
collapse {\it before} the global collapse of the cloud is completed
(Fig.\ \ref{fig:dense_turb}, {\it right panel}),
both because their densities are large enough that their free-fall time
is significantly shorter than that of the whole cloud, and because the
free-fall time of a flattened or elongated cloud may be much larger than
that of an approximately isotropic clump within it of the same volume
density \citep{Toala+12}. However, the turbulent velocities initially
induced in the clouds by the converging flows in the simulations of
\citet{VS+07} were observed to be small compared to the velocities that
develop due to the subsequent gravitational contraction, while in the
simulations of \citep{Banerjee+09}, the clumps with highest
internal velocity dispersions were those that had already formed
collapsed objects (``sink'' particles), although energy feedback from
the sinks was not included.

All of the above evidence suggests that the observed supersonic motions
in molecular clouds may have a significant, perhaps dominant, component
of infalling motions, with a (possibly subdominant) superposed random
(turbulent) component remaining from the initial stages of the cloud
\citep{BBB03, BP+11a, BP+11b}, and perhaps somewhat amplified by the
collapse \citep{VS+98}. In this scenario of {\it hierarchical
gravitational fragmentation}, the first structures that complete their
collapse are small-scale, high-amplitude density fluctuations that are
embedded within larger-scale, smaller amplitude ones, which complete
their collapse later \citep{VS+09}. The main role of the truly turbulent
(i.e., fully random) motions is to provide the nonlinear density
fluctuation seeds that will collapse locally after the global
contraction increases their density sufficiently for them to become
locally gravitationally unstable \citep{CB05}. Evidence for such
multi-scale collapse has recently begun to be observationally detected
\citep{Galvan+09, Schneider+10}.

\section{Summary and Conclusions} \label{sec:conclusions}

In this contribution, we have briefly reviewed the role of two
fundamental physical processes (net radiative cooling and self-gravity)
that shape the nature of interstellar turbulence. We first discussed the
effects of the net thermal effects that arise from several microphysical
processes in the ISM, such as the emission of radiation from various
ions, atoms, and molecules, which caries away thermal energy previously
stored in the particles by collisional excitation, thus cooling the gas,
and photoelectric production of energetic electrons off dust grains by
background stellar UV radiation, which heats the gas. The presence of
radiative heating and cooling implies in general that the gas behaves in
a non-isentropic (or non-adiabatic) way, and in particular it may
become {\it thermally unstable} in certain regimes of density and
temperature, where small (i.e., {\it linear}) perturbations can cause
runaway heating or cooling of the gas that only stops when the gas exits
that particular regime. This in turn causes the gas to avoid those 
unstable density and temperature ranges, and tend to settle in the 
stable ones, thus tending to segregate the gas into different phases of
different densities and/or temperatures. In classical models of the ISM,
only the stable phases were expected to exist.

We then discussed the interaction between trans- or supersonic
turbulence, which produces large (i.e., {\it nonlinear}) density and
velocity, and thermal instability (TI). We first discussed the
probability density function (PDF) of the density fluctuations, which
takes a lognormal form in isothermal regimes, and develops power-law
tails in polytropic (i.e., of the form $P \propto \rho^\gamef$) ones. We
then noted that, since turbulence is an inherently mixing phenomenon,
it opposes the segregating effect of thermal instability, causing the
production of gas parcels in the classically forbidden unstable regimes,
which may add up to nearly half the mass of the ISM, although the
density PDF in general still exhibits some multimodality due to the gas'
preference to settle in the stable regimes. The existence of gas in the
unstable ranges has been confirmed by numerous observational studies.

We next discussed the nature of the turbulence in the different ranges
of density and temperature of the gas, noting that in the diffuse
ionized regions, where the flow is transonic (i.e., with Mach numbers
$\Ms \sim 1$), the gas appears to behave in an essentially incompressible
way, exhibiting Kolmogorov scalings over many orders of magnitude in
length scale. However, in the neutral atomic regions, where the gas is
expected to be thermally unstable under the so-called {\it isobaric
criterion}, the flow is expected to exhibit large density and
temperature fluctuations, by up to factors $\sim 100$, thus being highly
fragmented. Numerical simulations of this process suggest that the gas
is transonic with respect to the warm diffuse component, but supersonic
with respect to the cold, dense one, although those supersonic motions
appear to correspond more to the velocity dispersion of the dense clumps
within the warm substrate, rather than to the internal velocity
dispersion within the clumps themselves.

Finally, we pointed out that large-scale compressions in the warm
neutral gas, which may be triggered by either random turbulent motions,
or by yet larger-scale instabilities, may nonlinearly induce the
formation of large regions of dense, cold gas; much larger, in
particular, than the most unstable scales of TI, which have sizes $\sim
0.1$ pc. These large clouds may easily be large enough to be {\it
gravitationally} unstable, and numerical simulations of their evolution
suggest that they rapidly engage in gravitational contraction. The
latter may in fact promote the formation of molecules, so that the
clouds are likely to become molecular only after they begin
contracting. In addition, the clouds are born internally turbulent by
the combined effect of TI and other dynamical instabilities, and the
resulting nonlinear density fluctuations (``clumps'') may themselves
become locally gravitationally unstable during the contraction of the
whole large-scale cloud. Because they are denser, they have shorter
free-fall times, and can complete their local collapses before the
global one does, thus producing a regime of {\it hierarchical gravitational
fragmentation}, with small-scale, short-timescale collapses occurring
within larger-scale, longer-timescale ones. It is thus quite
likely that the flow regime in the dense molecular clouds corresponds to
a dominant multi-scale gravitational contraction, with smaller-amplitude
random (``turbulent'') motions superposed on it. Interstellar
turbulence is seen to involve an extremely rich and complex
phenomenology, even more so than the already-fascinating regimes of
terrestrial turbulence.



\end{document}